# A framework for seizure detection using effective connectivity, graph theory and deep modular neural networks

Behnaz Akbarian, Abbas Erfanian*

Iran Neural Technology Research Centre, Department of Biomedical Engineering,
Iran University of Science and Technology (IUST), Iran, Tehran

## Abstract

*Objective*

The electrical characteristics of the EEG signals can be used for seizure detection. Statistical independence between different brain regions measures by functional brain connectivity (FBC). Specific directional effects can't consider by FBC and thus effective brain connectivity (EBC) is used to measure causal intervention between one neuronal region and the rest of the neuronal regions. Our main purpose is to provide a reliable automatic seizure detection approach.

*Methods*

In this study, three new methods are provided. Deep modular neural network (DMNN) is developed based on combination of various EBC classification results in the different frequencies. Another method is named "modular effective neural networks (MENN)". This method combines the classification results of the three different EBC in the specific frequency. "Modular frequency neural networks (MFNN)" is another method that combines the classification results of the specific EBC in the seven different frequencies.

*Results*

The mean accuracy of the MFNN are 97.14%, 98.53%, and 97.91% using directed transfer function, directed coherence, and generalized partial directed coherence, respectively. Using the MENN, the highest mean accuracy is 98.34%. Finally, DMNN has the highest mean accuracy which is equal to 99.43. To our best knowledge, the proposed method is a new method that provides the high accuracy in comparison to other studies which used MIT-CHB database.

*Conclusion and significance*

The knowledge of structure-function relationships between different areas of the brain is necessary for characterizing the underlying dynamics. Hence, features based on EBC can provide a reliable automatic seizure detection approach.

*Corresponding author.
*E-mail addresses*: akbarian_b@elec.iust.ac.ir (B. Akbarian), erfanian@iust.ac.ir (A. Erfanian)

# 1. Introduction

Epilepsy, a neurological disorder, is one of the most common neurological disease that can be specified by inconsistent seizures. A procedure to diagnose seizures is visual inspection of the electroencephalography (EEG). This work is very time-consuming and error-prone [1]. Therefore, in recent years, automatic seizure detection has attracted a lot of attention and various algorithms are presented based on frequency analysis [2-5], time analysis [4-6], time-frequency analysis [7-11], and nonlinear analysis [12, 13].

One of the methods for characterizing brain functions is functional brain connectivity (FBC). Functional connectivity is a statistical concept that measures statistical independence between different neuronal regions. Although, the functions of the brain regions are different, the brain is not considered as a collection of independently functioning nodes [14]. The structure-function relationship between different areas of the brain causes an amount of overlap between various areas [14]. Existence of overlap means that for a particular task, different nodes or areas of the brain may be involved, which this concept is known as functional integration. The functional integration of the healthy state is different from the pathological state such as epilepsy [14]. The seizure starts from an area and spreads to other regions of the brain that is called the seizure propagation, thus, several regions of the brain are activated in seizure event [15]. Therefore, the study of functional connection is the appropriate method for seizure detection. Functional connectivity can be estimated in a various way, such as correlation (COR), cross-correlation (xCOR), covariance (COV), coherence (COH), phase-locking value (PLV) and mutual information (MI). The limitation of FBC is that does not consider specific directional effects.

Effective brain connectivity (EBC) measures intervention between one neuronal region and the rest of the neuronal regions and considers directional effects [16]. The EBCs based on Granger causality (GC) are appropriate for EEG signals because the brain signals can be predicted with the help of previous information about the signal, or on the other hand, brain signals are casual [16]. Directed transfer function (DTF), direct DTF (dDTF), partial directed coherence (PDC), directed coherence (DC), generalized partial directed coherence (GPDC), full-frequency DTF (ffDTF) are examples of the effective connectivity based on GC.

Localizing the seizure focus is an important step for seizure surgery in refractory patients. The main purpose of some studies is to detect the propagation pattern across brain areas and identify seizure onset zone (SOZ) [18-25]. These studies show that the brain connectivity patterns can offer practical information about seizure propagation and thus enhance the accuracy of the pre-surgical evaluation in patients with refractory epilepsy.

Seizure detection is another application of FBC and EBC [26-36]. In [26, 27], total information outflow based on the DTF and PDC considered as feature and SVM is a classifier, respectively. In [28], PSI of the multichannel ECoG data is used to distinguish between seizure and interictal activity. In [29], the brain connectivity based on COH for the different bands is used for discriminating between psychogenic nonepileptic seizures (PNES) and focal epilepsy. Mean clustering coefficient and average weighted path length are used to describe properties and the topology of the network. The fourth-order information based on common spatial pattern (CSP) of the COH matrices are used as features and SVM applied as classifier. In [30], a modification of the cosine similarity index is used as functional connectivity. Several graph theory measures are used as features and Gaussian mixture model (GMM) is used for seizure detection. In [31], distinctions between theta and beta frequency bands of the epilepsy groups and healthy groups are investigated. Outflow information on the different frequencies in the beta and theta bands are calculated by using weighted PDC. This study shows that theta band variations and beta band variations are different in patterns. In [32], functional connectivity of the MEG for six frequency bands are computed and graph theory measures such as normalized clustering coefficient and path length are extracted as features. In [33], graph theory features based on several FBC and EBC considered as a feature vector to classify ECoG signals of rats before and after pentylenetetrazole (PTZ) injection. In [34-36], closed loop neurostimulators are provided for seizure suppression. In these systems, a seizure detection algorithm is required. In [34], PLV and in [35, 36] phase synchrony index are used to detect the onset seizure detection.

In this study, a novel method, deep modular neural networks (DMNN), is presented for seizure detection. In this method, information about different frequencies and different effective connectivity are combined with each other to achieve maximum accuracy. The combination of global and local graph theory measures are used as features. These features represent the network and nodal properties of the EBCs. In the event of a seizure, in addition to the importance of information in an individual electrode, the connection of the electrode to the other electrodes is also important. Therefore, a combination of both features has been used. Autoencoder (AE) is used for feature mapping and dimension reduction. At first, these features are not suitable for classification, but when these features map to new space using AE they become appropriate for seizure detection. Then, features are fed to softmax classifier for seizure detection and results of classifiers are combined to achieve different modular structures. The results are very encouraging with 99.43% mean accuracy for all patients. The remainder of this paper is devoted to the description of the method in Section 2, result in Section 3, discussion in Section 4, and conclusion in Section 5.

## 2. Methods

In Fig. 1, the schematic of the proposed method is shown. As can be seen in Fig. 1, at first, 23-channel EEG signals are split into 1.2 s non-overlapping segments. Then, effective connectivity are calculated from these segmented signals. Global (1×number of segments), and local (number of channels × number of segments) graph theory measures extracted from the EBC matrixes (23×23). These features are fed to AE for feature mapping and dimension reduction. The new features are used for seizure detection by softmax classifier. After all, the results of the classifiers are combined using different combing rule algorithms that include modular frequency neural networks (MFNN), modular effective neural networks (MENN) and deep modular neural networks (DMNN). Each section of the procedure is described below.

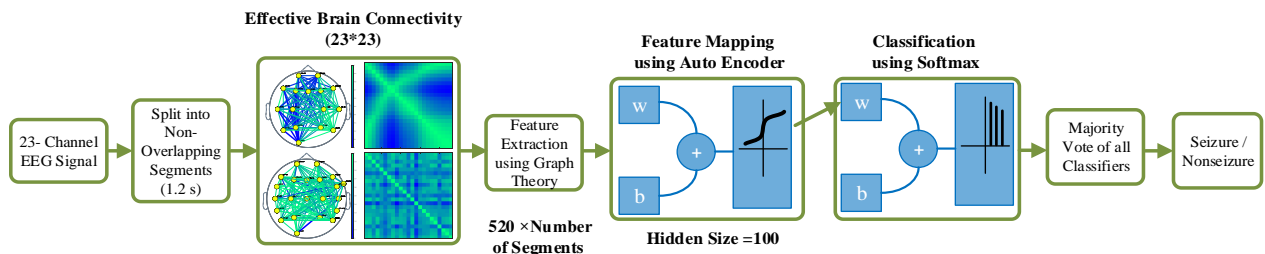

**Fig. 1.** Schematic of the proposed method.

## 2.1. Effective Brain Connectivity (EBC)

Effective connectivity attempts to extract networks of causal influences of one neural element over another. Actually, interaction between neurons should be modeled for EBC estimation. Various techniques for extracting effective connectivity have been pursued. Some of these methods are based on interpretations or adaptations of the concept of GC [37]. The multivariate autoregressive models (MVAR) can provide a temporal model from effect of on brain region on another. The MVAR is a model which indicates the multivariate signals as follows:

$$\begin{bmatrix} x_1(n) \\ \vdots \\ x_N(n) \end{bmatrix} = \sum_{r=1}^{P} A_r \begin{bmatrix} x_1(n-r) \\ \vdots \\ x_N(n-r) \end{bmatrix} + \begin{bmatrix} w_1(n) \\ \vdots \\ w_N(n) \end{bmatrix} \quad (1)$$

where $P$ is the model order. In this work, Akaike Information Criterion (AIC) is used to select the model order. $A_r$ is MVAR parameters that are estimated by multichannel Yule-Walker. $w = [w_1(n), \dots, w_N(n)]$ is uncorrelated white noise at time $n$, $x(n)$ is signal at time $n$, $x(n-r)$ is the past value of the $x(n)$, and $N$ indicates the number of signals (channels).

The casual relation between signals is examined in the spectral domain by calculation the Fourier transformation of Eq. (1). The transfer matrix of the MVAR model $H(f)$, and power spectral density (PSD) matrix, $S(f)$ are estimated as follows:

$$H(f) = \bar{A}^{-1}(f) = \big(I - A(f)\big)^{-1} \quad (2)$$

$$S(f) = H(f) \Sigma H^H(f) \quad (3)$$

$$A(f) = \sum_{r=1}^{P} A_r z^{-r} \big|_{z = e^{-i 2\pi f}} \quad (4)$$

$$A_r = \begin{bmatrix} a_{11}(r) & \cdots & a_{1N}(r) \\ \vdots & \ddots & \vdots \\ a_{N1}(r) & \cdots & a_{NN}(r) \end{bmatrix} \quad (5)$$

where $\Sigma$ is the noise covariance matrix. The $a_{ij}(r)$ shows the linear interaction effect of $x_j(n-r)$ onto $x_i(n)$. Then, the DTF, DC and GPDC are defined as follows [38]:

$$DTF_{ij}(f) = H_{ij}(f) / \left( \sqrt{\sum_{j=1}^{N} |H_{ij}(f)|^2} \right) \quad (6)$$

$$DC_{ij}(f) = H_{ij}(f) / \sqrt{S_{ij}(f)} \quad (7)$$

$$GPDC_{ij}(f) = \frac{{1}/{\sigma_i} |\bar{A}_{ij}(f)|}{\sqrt{\sum_k {1}/{\sigma_k^2} |\bar{A}_{kj}(f)|^2}} \quad (8)$$

where $\sigma_i^2$ is the variance and $f$ is the frequency that is set to values 1, 4, 8, 12, 16, 20, and 24 Hz.

## 2.2. Graph Theory Measures

A graph consists of two main parts, including nodes (neurons or brain regions) and links (synapses or pathways, or statistical dependencies between neural elements). Unlike binary graphs that only indicate there is a connection between two nodes or not (0=absence, and 1=existence), the weighted graphs show the strength of the connection between the multiple nodes. Graph theory is a method for the quantification and analysis graph. Graph theory can propose important new insights into the structure and function of networked brain systems, including their architecture, evolution, development, and clinical disorders [39]. Graph measures can be subdivided into two broad classes:

Global (network) measures mention global properties of a graph and, therefore, consist of a single value for each graph.

Local (nodal) measures refer to properties of the nodes of a graph and, therefore, represented as a vector with length equal to the number of the nodes.

### 2.2.1. Transitivity

The ratio of triangles to triplets in the network is Transitivity. Transitivity is a global measure [40, 41]. Transitivity can be defined as follows:

$$\text{Transitivity} = (3 \times \text{number of triangles in the network})/\text{number of connected triples of vertices} \tag{9}$$

### 2.2.2. Node strength

The sum of weights of all links connected to a node is node strength. The in-strength is the sum of inward link weights, and the out-strength is the sum of outward link weights. All measures are local measure. Therefore, node strength is the sum of all rows, and columns of a weighted connectivity matrix. In-strength (out-strength) is the sum of all columns (rows) of a weighted connectivity matrix [42].

### 2.2.3. Participation coefficient

Participation coefficient is a nodal measure which measures the diversity of the node's connections across communities. When participation coefficient of a node is high, this means that corresponding node is connected to many communities [41, 43].

### 2.2.4. Within-module degree z-score

The within-module degree z-score is a local measure. It indicates which a node is connected to the other nodes in the same community [41, 43]. This measure is divided into two categories:

Within-module in-z-score: To calculate it only the contribution of in-path lengths should be considered. Within-module out-z-score: To calculate it only the contribution of out-path lengths should be considered. The z-score is calculated as Eq. (10)

$$Z_i = (K_i - K_{S_i})/\sigma_{S_i} \tag{10}$$

where $S_i$, $K_i$, $K_{S_i}$ and $\sigma_{S_i}$ are the community that the node belongs to, the degree of the node in the community $S_i$, the average degree of all nodes in the community $S_i$, and standard deviation of the nodes degree within the community $S_i$, respectively.

### 2.2.5. Modularity

The degree that the network may be subdivided into clearly specified communities (i.e. groups, sub graphs or modules) is measured by the modularity. Modularity is a global statistic measure. Modularity is calculated according to the Eq. (11):

$$Modularity = \frac{1}{l}\sum_{ij}\left[A_{ij} - \frac{k_i k_j}{l}\right]\delta_{ij} \qquad (11)$$

where $A_{ij}$, $k_i$ ($k_j$) and $l$ are the connectivity matrix, degree of the node $i$ ($j$), and the number of edges in the graph, respectively. If two nodes belong to the same community $\delta_{ij} = 1$, otherwise is equal to zero [41, 43].

### 2.2.6. Global efficiency and local efficiency

The average of inverse shortest path length from a node to all other nodes is the global efficiency. In other words, it is inversely associated with the characteristic path length. When the global efficiency computed on the neighborhood of the node it is defined as the local efficiency, and is associated with the clustering coefficient [41, 44].

### 2.2.7. Mean global diffusion efficiency

The diffusion efficiency is measured by the expected number of steps that takes a random walker starting at node $i$ to arrive at node $j$ for the first time [45].

### 2.2.8. Clustering coefficient

The average "intensity" (geometric mean) of all triangles associated with each node is defined as the clustering coefficient. In other words, the clustering coefficient is the ratio between triangles which present around a node and maximum number of triangles. It could be constituted around node and is a local measure. In a directed graph, the total number of triangles is calculated as $d_{in} \times d_{out} - d_{ii}$. Where $d_{in}$, $d_{out}$, and $d_{ii}$, are in-degree of a node, out-degree of a node, and number of links that cannot constitute triangles [41, 46].

### 2.2.9. Characteristic path length and related statistics

The average shortest path length between all pairs of nodes in the network is the network characteristic path length; on the other hand, it is average of the path lengths of all nodes. It is a global measure [41, 46]. The maximal path length (distance) between a certain node and any other node in the network is the eccentricity and is a local measure. Minimal eccentricity is the radius and the diameter is the maximal eccentricity. Radius and diameter are global measures [46].

### 2.2.10. Assortativity coefficient

The assortativity coefficient is a global measures that quantifies correlation coefficient between the strengths of all nodes on two opposite ends of a link. Positive assortativity coefficient shows nodes tend to connect to other nodes with the same or similar strength. It is calculated as Eq. (12):

$$r = \frac{l^{-1}\sum_{i,j\in L} k_i k_j - \left[\left[l^{-1}\sum_{i,j\in L}\frac{1}{2}(k_i + k_j)\right]^2\right]}{l^{-1}\sum_{i,j\in L}\frac{1}{2}(k_i^2 + k_j^2) - \left[\left[l^{-1}\sum_{i,j\in L}\frac{1}{2}(k_i + k_j)\right]^2\right]} \tag{12}$$

where $k_i$ ($k_j$) and $l$ are degree of the node $i$ ($j$), and the number of edges in the graph, respectively. The assortativity coefficient can be calculated for four modes such as out-strength/in-strength correlation, in-strength/out-strength correlation, out-strength/out-strength correlation, in-strength/in-strength correlation [41, 47].

*2.3. Single Layer Autoencoder*

Deep learning is an appropriate method of machine learning. AE is a symmetrical neural network which can use to extract features in an unsupervised manner [48]. AE has encoding layer and decoding layer. Feature extraction is done by minimizing the reconstruction error between the input data at the encoding layer, and its reconstruction at the decoding layer [49]. The features generated by the AE can fed to any other classifier, and be used for classification. In general, the main purpose of the AE is to learn a representation (encoding) for a dataset, and is generally used to reduce the dimension [49]. In this study, the feature vector with a dimension of 512 has been reduced to a feature vector with dimension of 100 using AE. Therefore, the final feature vector is a compressed representation of the original feature vector input. The main characteristic of an AE is the ability to extract features from a large number of unlabeled data [48]. In this study, the developed AE contains an encoder for learning first-order features of the EEG feature vector input.

*2.4. Modularity Classification*

Modularity classification is a breakdown of the classification work into simpler tasks to improve their performance [50, 51]. A softmax layer is trained as a simple classifier for seizure detection, and results of the multiple softmax classifiers are used to final decision. Three different modularity classification methods have been used which are MFNN, MENN, and DMNN.

The seizure progression includes multiple frequencies. Most seizure activities are in the frequency range 0.5 Hz to 30 Hz [12]. In this paper, seven frequencies with step of 4 Hz apart from each other are selected as representatives of the frequencies between 0 and 30 Hz. In MFNN approach, different effective connectivity such as DTF, DC, and GPDC are calculated in various frequencies including 1, 4, 8, 12, 16, 20, and 24 Hz. Then, graph theory measures are calculated as features. These features fed to AE for future reduction. The new features in each frequency are classified with a softmax classifier. Thus, there are seven softmax classifier that the results of the seven soft max classifiers are combined together for final decision by majority voting. MFNN method is provided for considering different frequency information, and is shown in Fig. 2.

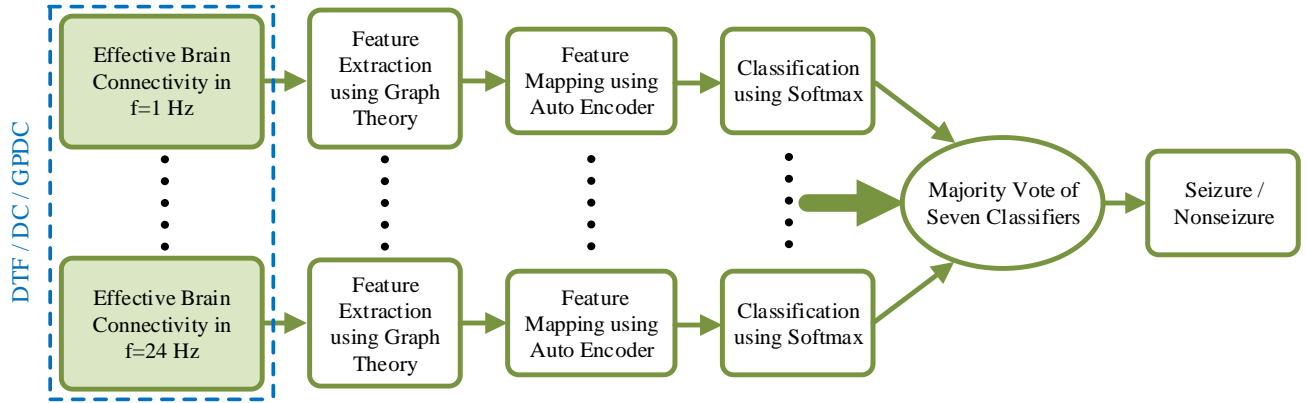

**Fig. 2.** Schematic of the modular frequency neural network (MFNN) method.

DTF, DC, and GPDC are used to investigate the relationships between different channels. In MENN approach, DTF, DC, and GPDC are calculated in one of the frequencies 1, 4, 8, 12, 16, 20 , and 24 Hz; hence, there are three matrices 23×23 in the same frequency. After calculation of the graph theory measures and feature mapping by AE, the results of the three softmax classifiers based on different effective are combined with the majority vote. The MENN method is shown in Fig. 3.

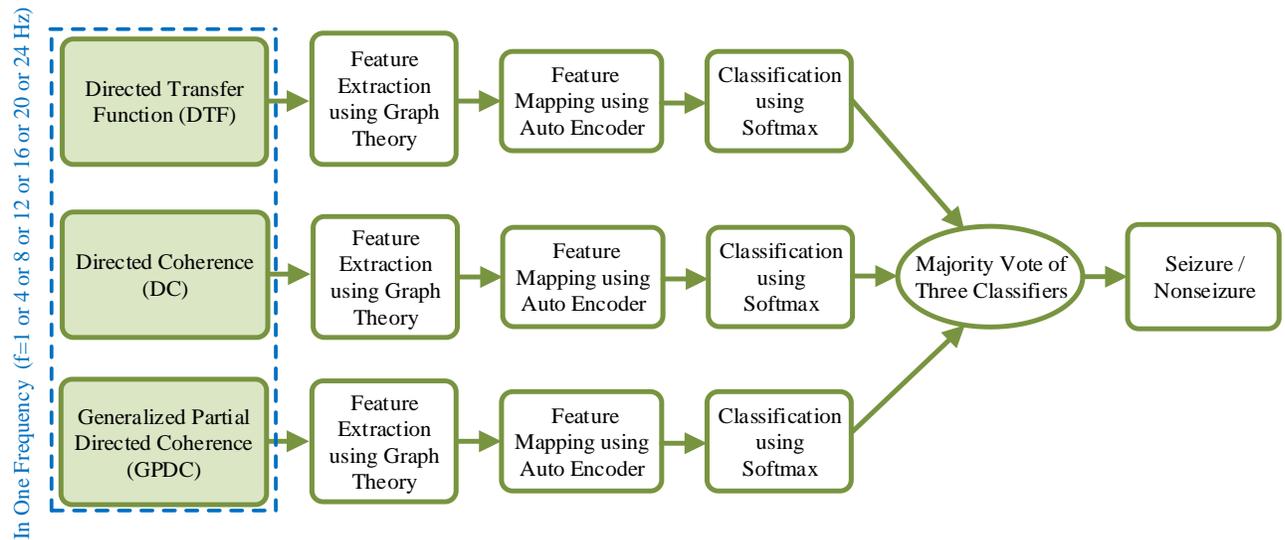

**Fig. 3.** Schematic of the modular effective neural network (MENN) method.

Finally, the method that is called DMNN is provided. In this method, information about different frequencies and different effective connectivity are combined together. In the step 1, DTF, DC, and GPDC are calculated in seven frequencies (1, 4, 8, 12, 16, 20 and 24 Hz); thus, there are 7 matrices 23×23 for each of them. Then, the graph theory measures are used as features and AE implied for feature extraction from original features. The results of the feature classification for each frequency in the specific EBC are combined with the majority vote of the seven softmax classifiers in the step 1. In the step 2, three obtained results of the step 1 are combined with the majority vote. The DMNN method is shown in Fig. 4.

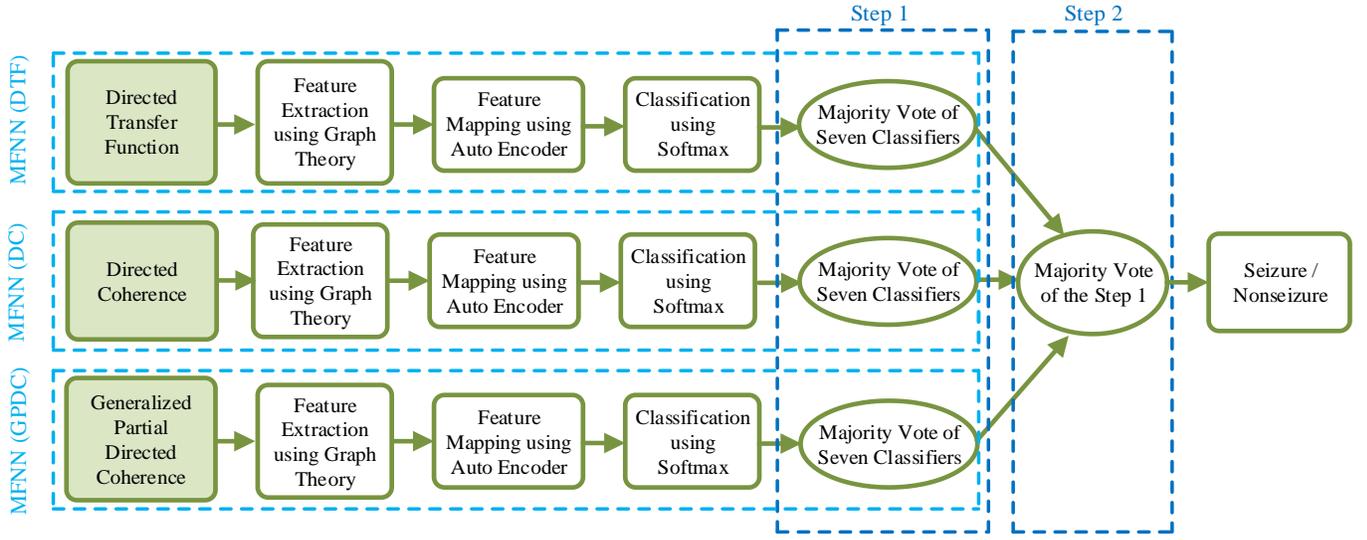

**Fig. 4.** Schematic of deep modular neural network (DMNN) method.

### 2.5. Dataset

To validate the performance of the proposed method, the CHB-MIT long-term scalp EEG database has been used [52, 53]. The used dataset was provided by Children's Hospital in Boston (CHB), and the Massachusetts Institute of Technology (MIT), which includes 23 pediatric subjects (5 males, and 17 females, the subject 1 and 21 are the same patient thus 23 subjects are 22 individual patients) with intractable seizures. A summary of patient's information is presented in Table 1. This information includes age, gender, number of seizure events, and seizure duration.

The sampling rate of EEG signals was 256 Hz, quantized with a 16 bit analog to digital converter, recorded by 23-channels scalp EEG based on 10-20 international standard acquisition system. We divided signals into segments with the length of 307 samples (1.2 seconds). For preprocessing stage, the signals were normalized, and filtered through a pass filter with the frequency range from 0.5 Hz to 25 Hz. For seizure detection, two types of signals were defined according to previous research. EEG segments labeled as 'seizure' and 'nonseizure'

### 3. Result

The performance of the proposed method is checked with the classification accuracy (CA). CA is proportion of the correctly classified EEGs out of the total number of EEGs. By using this criterion, the classification behavior can be estimated on the extracted features. The definition of this parameter is as Eq. (13):

$$CA = \frac{TP + TN}{TP + FN + TN + FP} \times 100\% \tag{13}$$

where, TP= True Positive, FN= False Negative, TN= True Negative, FP= False Positive.

**Table 1.**

Summary of CHB-MIT Database.

| Patient ID | Gender, Age | # of seizures | Seizure Duration (sec) | # of Segment for each class |
|---|---|---|---|---|
| **chb01** | F, 11 | 7 | 442 | 368 |
| **chb02** | M, 11 | 3 | 172 | 143 |
| **chb03** | F, 14 | 7 | 402 | 335 |
| **chb04** | M, 22 | 4 | 378 | 315 |
| **chb05** | F, 7 | 5 | 558 | 465 |
| **chb06** | F, 1.5 | 10 | 153 | 127 |
| **chb07** | F, 14.5 | 3 | 325 | 270 |
| **chb08** | M, 3.5 | 5 | 919 | 765 |
| **chb09** | F, 10 | 4 | 276 | 230 |
| **chb10** | M, 3 | 7 | 447 | 372 |
| **chb11** | F, 12 | 3 | 806 | 671 |
| **chb12** | F, 2 | 40 | 1140 | 950 |
| **chb13** | F, 3 | 12 | 535 | 445 |
| **chb14** | F, 9 | 8 | 166 | 138 |
| **chb15** | M, 16 | 20 | 1992 | 1660 |
| **chb16** | F, 7 | 10 | 84 | 70 |
| **chb17** | F, 12 | 3 | 293 | 244 |
| **chb18** | F, 18 | 6 | 317 | 264 |
| **chb19** | F, 19 | 3 | 236 | 196 |
| **chb20** | F, 6 | 8 | 294 | 245 |
| **chb21** | F, 13 | 4 | 199 | 165 |
| **chb22** | F, 9 | 3 | 204 | 170 |
| **chb23** | F, 6 | 7 | 424 | 353 |

## 3.1. Effective Brain Connectivity

The type of healthy network and epileptic network is one of the points that has been considered. Several types of networks can be regarded, which include: Ordered (regular or lattice-like) networks, random networks, small-world networks [54].

Fig. 5 and Fig. 6 show typical network topology of the DTF, DC, and GPDC for the nonseizure, and seizure at $f = 1$ and $f = 24$ Hz, respectively. As can be seen, the interaction, and information flow in seizure is different from nonseizure. During seizure activity the neuronal network changes in direction of a small-word network [55, 56]. The nonseizure neuronal network has a more random configuration than seizure topology. The number of the brain regions and their information flow between them are increased during seizures.

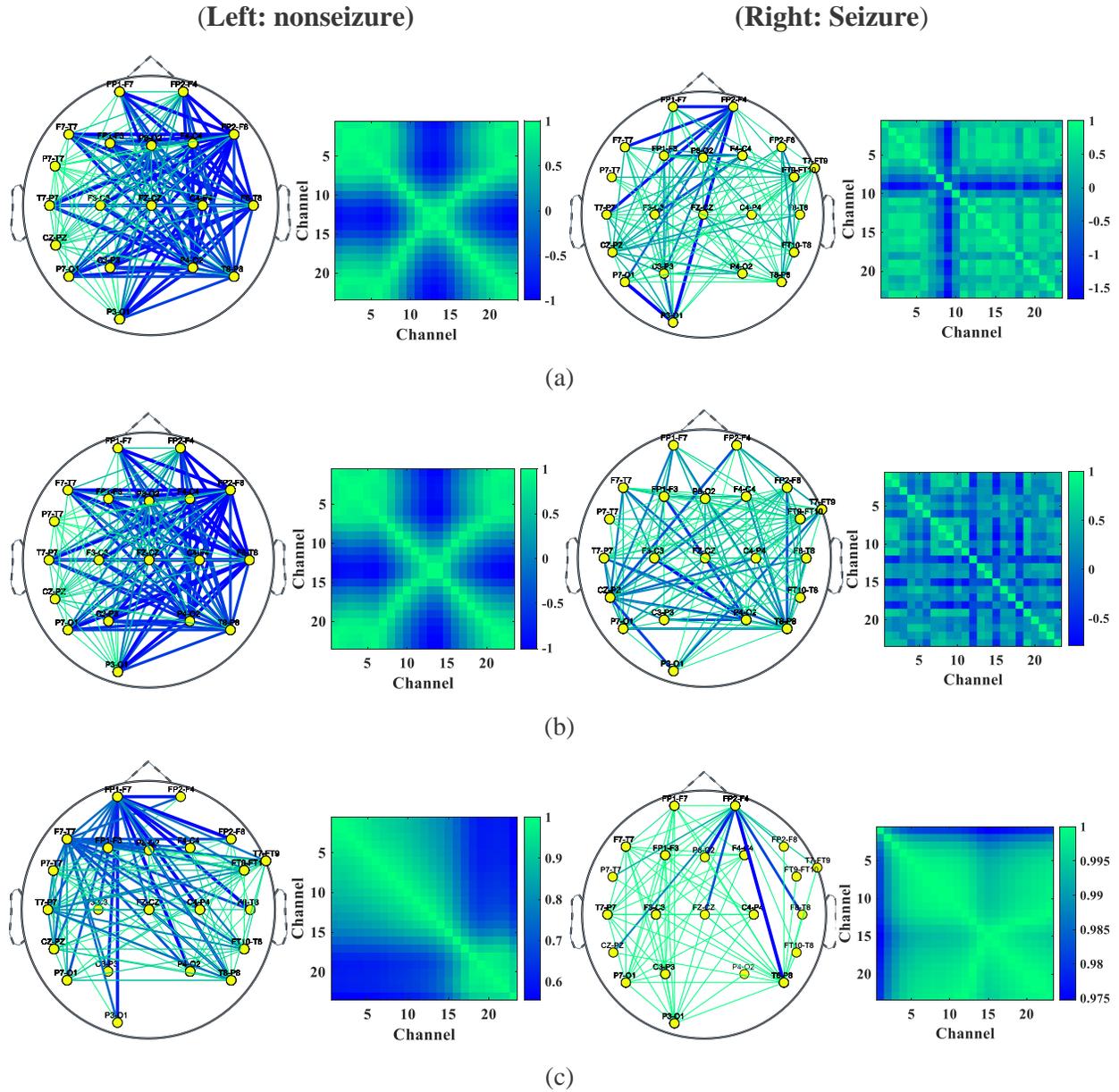

**Fig 5.** Left: Brain connectivity and effective connectivity networks for nonseizure. Right: Brain connectivity and effective connectivity networks for seizure. (a) DTF ($f = 1$), (b) DC ($f = 1$) and (c) GPDC ($f = 1$). (The figures are related to chb15).

**(Left: nonseizure)** **(Right: Seizure)**

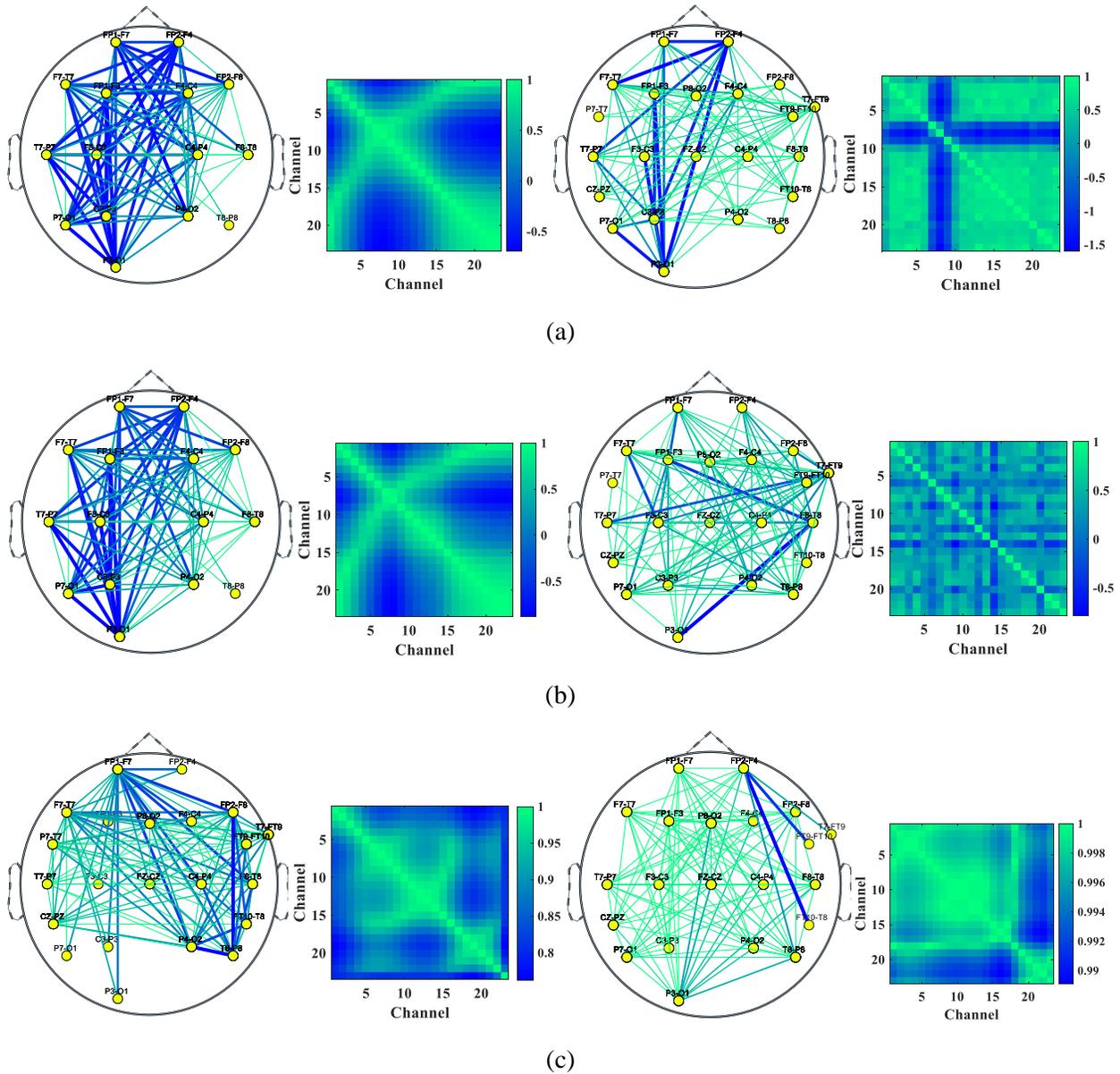

**Fig 6.** Left: Brain connectivity and effective connectivity networks for nonseizure. Right: Brain connectivity and effective connectivity networks for seizure. (a) DTF ($f = 24$), (b) DC ($f = 24$) and (c) GPDC ($f = 24$). (The figures are related to chb15).

Mean clustering coefficient can indicates the nature of a network. Decrease in the mean clustering coefficient shows increase in connectivity between different neural units [57]. The small-word network has mean clustering coefficient higher than random networks. Fig. 7 shows bar histograms of the average mean clustering coefficient based on DTF and DC. The mean clustering coefficient based on DC and DTF are 1.2241 and 1.0825 for nonseizure also 1.1849 and 0.9442 for seizure. As it can be seen mean clustering coefficient of the seizure is lower than nonseizure. Therefore in seizure, connectivity increases. In other words, many different brain areas are activated simultaneously.

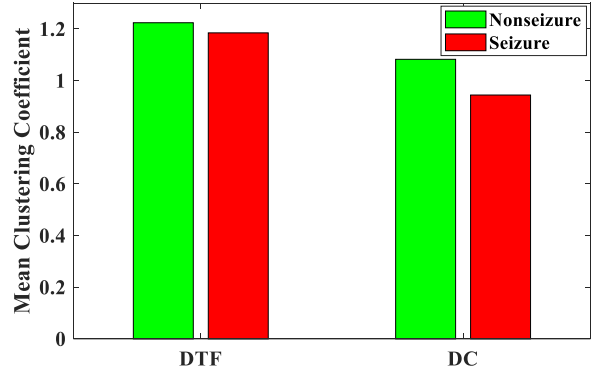

**Fig. 7.** Bar histograms of the average mean clustering coefficient.

The $K$-core is a criterion in order to investigate and visualize properties of the effective networks [58]. Fig 8 shows k-core decomposition ($K = 6$) of the seizure and nonseizure networks. As shown in Table 2, size of $K$-core based on DTF, DC and GPDC are 15, 13, and 14 in seizure and 10, 10, and 5 for nonseizure, respectively. The size of $K$-core of the seizure is more than nonseizure. This indicates that the degree of connectivity among the channels in seizure is more than nonseizure or number of brain regions that are associated with each other is increasing.

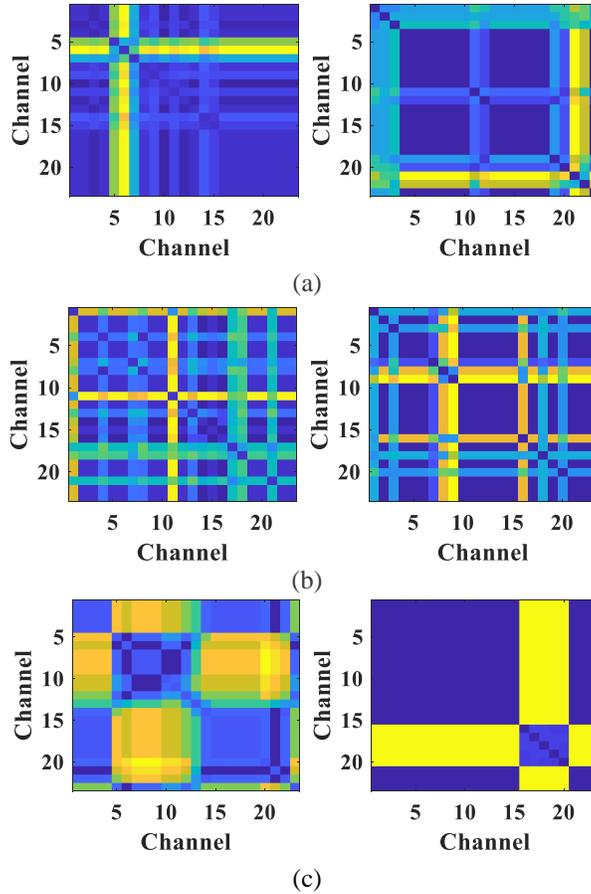

**Fig. 8.** Left: K-core decomposition ($K = 6$) of the seizure and right: K-core decomposition ($K = 6$) of the nonseizure networks. (a) DTF ($f = 1 + f = 2 + \cdots + f = 7$), (b) DC ($f = 1 + f = 2 + \cdots + f = 7$), and (c) GPDC ($f = 1 + f = 2 + \cdots + f = 7$), respectively.

**Table 2.**

Size of K-core for seizure and nonseizure.

| Size of K-core | DTF | DC | GPDC |
|---|---|---|---|
| **Seizure** | 15 | 13 | 14 |
| **Nonseizure** | 10 | 10 | 5 |

These results show that the nature of networks in seizure, and nonseizure are different with each other, hence EBC is appropriate method for seizure detection.

*3.2. Feature mapping*

In this work, an AE is used to reduce the dimension of the extracted features. The transfer function of the encoder, and decoder are logistic sigmoid function ('logsig'), and linear transfer function ('purelin'), respectively. Loss function used for training is mean squared error function ('msesparse'). The input is a 520-by-number of segments matrix defining 520 attributes. The size of hidden representation of the autoencoder is 100 neurons. Thus 520 original features map to 100 new features. Production of the new features by the AE is shown in Fig. 9. New features in the training are made in such a way that the reconstruction loss is minimized [59].

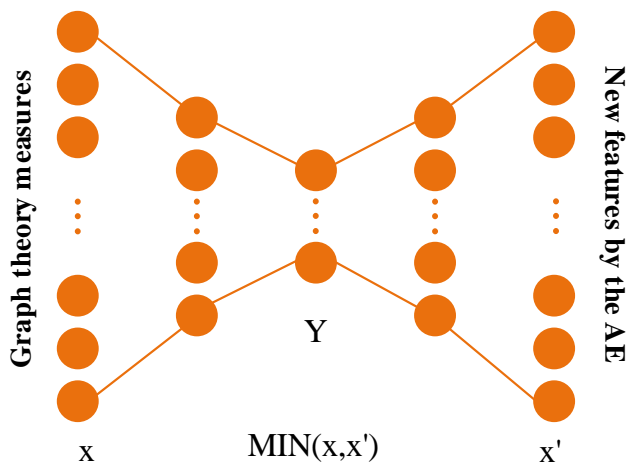

**Fig 9.** Schematic of the AE for extraction of the features.

Fig. 10 shows 520 original features, and 100 extracted features by the AE. When high dimensional space (520) reduce into less dimension new space (100) by using AE, the separation ability of the features increases. The t-Distributed Stochastic Neighbor Embedding (t-SNE) method has been used to indicate original and new features on three dimensions [60]. The t-SNE is a method for visualizing high-dimensional data by giving each data point a location in a two or three-dimensional map. As shown in Fig. 10, the features of the AE have a high power to separate classes.

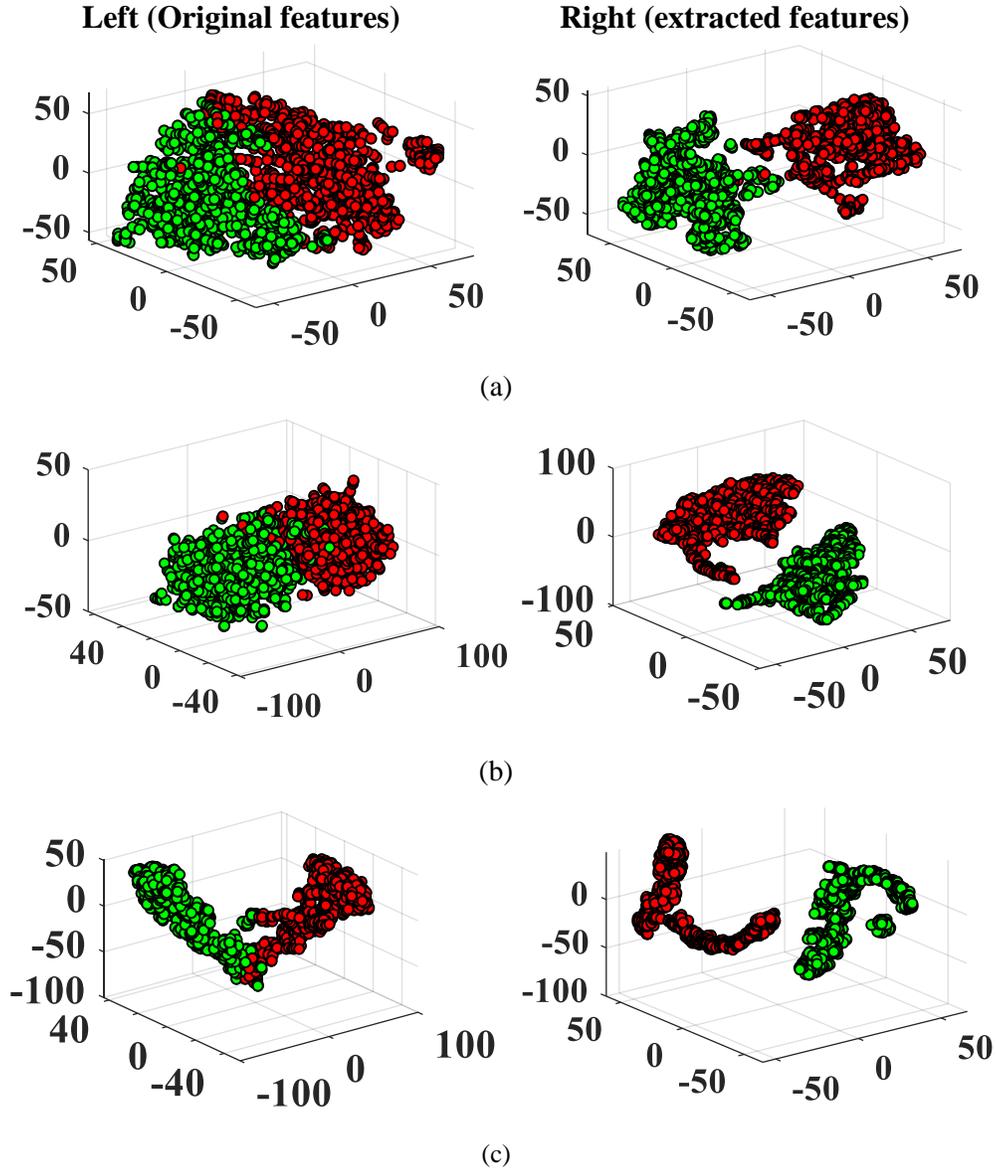

**Fig. 10.** The ability to separate 520 original features (Left) and 100 extracted features by the AE (Right) for chb18. (a) Features based on DTF, (b) Features based on DC, (c) Features based on GPDC.

## 3.3. Classification

Fig. 11 summarizes the classification results for effective connectivity in the each frequency and MFNN method. The highest mean accuracy in one frequency are 93.2 ($f = 20$), 94.73 ($f = 16$) and 96.56 ($f = 24$) for DTF, DC and GPDC, respectively. The mean accuracy of the MFNN method is 97.14, 98.53 and 97.91 based on DTF, DC, GPDC, respectively. That is shown when the information of different frequency are combined together using MFNN method. The classification performance has been improved.

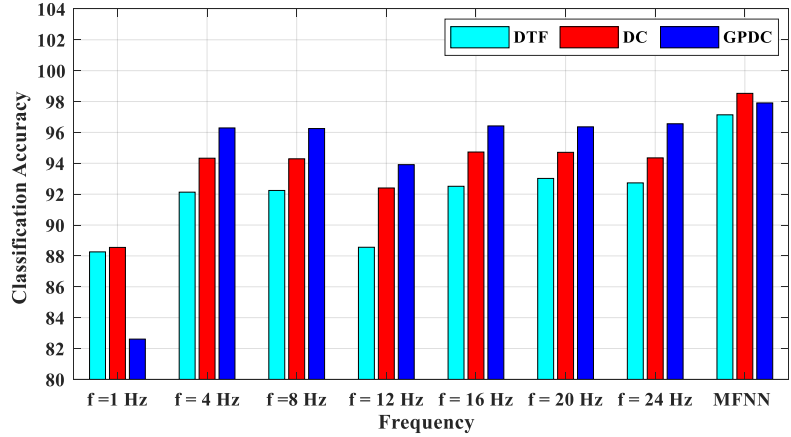

**Fig. 11.** Bar histograms of the mean classification accuracy for all patients in the different frequency using MFNN method.

Fig. 12 indicates the classification results for MENN method. Using the MENN method, the highest mean accuracy is related to the frequency 20 Hz (98.34%), which is increased compared to the MFNN.

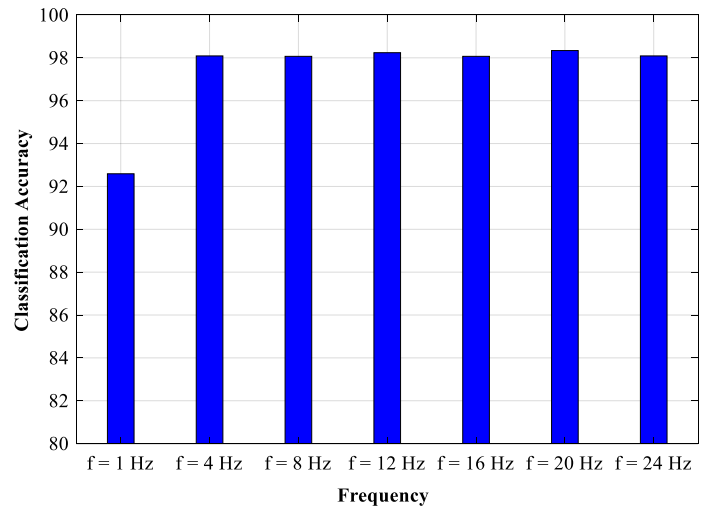

**Fig. 12.** Bar histograms of the mean classification accuracy for all patients in the different frequency using MENN method.

Table 3 summarizes the classification results for DMNN method. In the DMNN method, information of the all effective brain connectivity (DTF, DC, GPDC) and information of the all frequencies (1, 4, 8, 12, 16, 20, and 24 Hz) are used for classification. Finally, the DMNN method has the highest mean accuracy which is equal to 99.43% (maximum=99.99±0.01 for chb01 and minimum 93.38±5.11 for chb16). Fig. 13 shows results of different methods which the DMNN method has the highest accuracy.

**Table 3.**

Classification accuracy of the deep modular neural network (DMNN) method (mean±std).

| Patient ID | chb01 | Chb02 | chb03 | chb04 | chb05 | chb06 |
|---|---|---|---|---|---|---|
| Accuracy (mean±std) | 99.99±0.01 | 99.42±0.67 | 99.86±0.21 | 99.91±0.18 | 99.98±0.06 | 98.55±1.77 |
| **Patient ID** | **chb07** | **Chb08** | **chb09** | **Chb10** | **Chb11** | **Chb12** |
| Accuracy (mean±std) | 99.62±0.34 | 99.97±0.08 | 99.93±0.17 | 99.95±0.11 | 99.98±0.05 | 99.98±0.03 |
| **Patient ID** | **Chb13** | **Chb14** | **Chb15** | **Chb16** | **Chb17** | **Chb18** |
| Accuracy (mean±std) | 99.88±0.20 | 99.50±0.57 | 99.99±0.02 | 93.38±5.11 | 99.29±0.64 | 99.97±0.02 |
| **Patient ID** | **Chb19** | **Chb20** | **Chb21** | **Chb22** | **Chb23** | **Mean** |
| Accuracy (mean±std) | 99.26±0.67 | 99.2±0.62 | 99.86±0.31 | 99.38±0.93 | 99.96±0.11 | 99.43 |

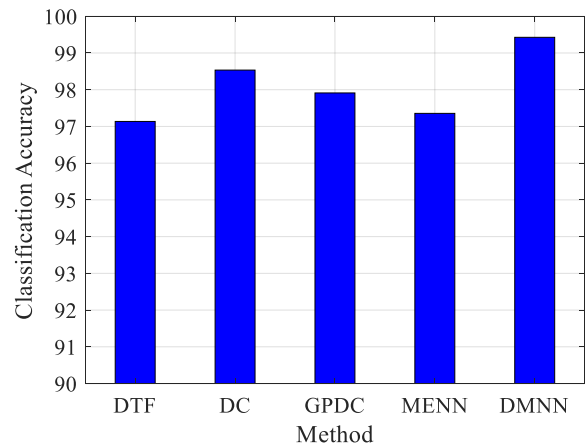

**Fig. 13.** Bar histograms of the mean classification accuracy using different methods.

## 4. Discussion

The analysis of brain signals is divided into two categories: single-channel analysis and multi-channel analysis. Multi-channel records provide a large amount of data and information. Single-channel approaches ignore structure-function relationships between different areas of the brain, while multi-channel methods such as EBC uses the information of the all channels. The knowledge of these structure-function relationships is necessary for characterizing the underlying dynamics. It can be noticed from Figs. 5 and 6 that the structure-function relationships increase in the seizure state.

In recent years, the development of the reliable automatic seizure detection approach, suitable for all patients, has on attention. The EEG signals are different in nature among patients. The nature of electromagnetic signals is dynamic and nonstationarity and seizure progression includes multiple frequencies. Most seizure activities are in the frequency range 0.5 Hz to 30 Hz [12]. In [61, 26, 33], connectivity is calculated at different frequencies, and eventually the connectivity are averaged over all frequencies and used as the final connectivity. In [19], the frequency that includes the highest power associated with seizure activity is selected using the Morlet wavelet transform. In [21, 22, 27, 29] by a band pass filter, the signal is divided into sub-frequency ranges, and in each subband, final connectivity is calculated by summing up connectivity values at different frequencies. With regards to these studies, in this study the MFNN method (Fig. 2) is used to combine information about different frequencies.

In addition to the importance of frequency, there are several effective connectivity to investigate the relationship between different channels. In [26, 33, 61] DTF, in [27, 33] PDC, and in [33] DC were used to seizure detection. In some studies, these effective connectivity were compared with each other, for example, in [61], the performance of the DTF and PDC is compared with each other, and it has been clearly shown that the PDC is more accurate than the DTF. In view of the fact that our main aim is to provide a framework that is highly accurate for a large number of patients, the results of the classification of the three different effective connectivity are combined with the MENN method (Fig. 3).

Finally, we have provided a method that is called DMNN (Fig. 4). In this method, information about different frequencies and different effective connectivity are combined together to form a classifier that is good for all patients. The results of the DMNN method indicates, feature extraction based on various EBC at the different frequencies. This method is a robust accurate seizure detection for all patients with a short period of time (1.2 s).

One of the graph theory measures is mean clustering coefficient which is an important parameter to figure out network topology. Network topology of seizure is ordered while network topology of nonseizure is random. Thus, it can be noticed from Fig. 7 that nonseizure has mean clustering coefficient higher than seizure. Other graph theory measures is the size of K-core that specifies the number of connections in a network. In the seizure state, extreme synchronization of large neuronal populations occurs leading to more brain regions are connected to each other, therefore, the size of K-core of the seizure is more than nonseizure. The variation of the K-core can be seen in Fig. 8 and Table 2.

The graph theory is a method for visualization and quantification of the EBCs and divided into two groups (global and local). In the event of a seizure, in addition to the importance of information in an individual electrode the connection of the electrode with other electrodes, network communication, is also important. Therefore, in this study a combination of both features global and nodal have been used. It can be seen in Fig. 10, the 520 original features can't separate two class properly. To solve this issue, AE is used for feature extraction from original features and the improvement of the separation ability.

The proposed method is also compared with the existing methods in Table 4. In Table 4, the highest value of achieved averaged CA is 99.43 using DMNN. It can be seen in Table 4 the proposed method is a new method that provides the high accuracy in comparison to other studies which used MIT-CHB database.

The restriction of the proposed method may be needs a large amount of data and consumes a lot of time in training phase. Recording a large amount of data for training may not be possible in a real-world. But considering this point that network topology in seizure and nonseizure are different, we decide to explore a way for pre-training a neural network by specific patterns of the network topology. This is known that seizure affects distinct brain regions in different patients. Thus, at first, we can use a method for SOZ and channel selection. These works can solve the limitations of the proposed procedure.

**Table 4.**

A comparison of the results obtained by the proposed method and others' methods.

| Authors, reference, and year | Average AC | # of penitent | # of feature | Block Duration | methods |
|---|---|---|---|---|---|
| Ahmad et al., [62] (2017) | 91.40% | 24 | 165 | 512 | Scattering transform, information fusion, Threshold |
| Ibrahim and Majzoub, [63] (2017) | 94.50% | 10 | 138 | 512 | Discrete wavelet transform (DWT), Shannon entropy, standard deviation, K-nearest neighbors (KNN) |
| Zhu et al., [64] (2017) | 88.90% | 23 | 23 | 512 | permutation entropy (PE) discriminant analysis (DA) |
| Xiang et al. [65] (2015) | 98.31% | 23 | # of selected channels | 256 | Fuzzy entropy, SVM |
| Shoeb et al. [52] (2004) | 94.24% | 23 | 84 | 256 | DWT, Energy SVM |
| This work | 97.14% | 23 | 7×100 | 307 | MFNN (DTF) |
| | 98.53% | | 7×100 | | MFNN (DC) |
| | 97.91% | | 7×100 | | MFNN (GPDC) |
| | 98.34% | | 3×100 | | MENN |
| | 99.43% | | Step 1 (7×100), Step 2 (3×100) | | DMNN |

## 5. Conclusions

In this study, a novel epileptic seizure detection approach has been proposed based on the combination of the estimated global and nodal theory measures from the different EBC in the different frequencies. AE is used to extract new features from original features. These extracted features are found very operative to differentiate seizure. The performance of the proposed method is evaluated using MIT-CHB database. In this study, three different structure is developed: DMNN, MENN, and MFNN. DMNN method is developed based on an AE to quantify global and nodal graph theory measures based on various EBC in the different frequencies to deep feature extraction. MENN is the method that combines features of the different EBC in the specific frequency and MFNN is the method that combines specific EBC in the different frequency. It was observed that the CA increase with the DMNN method. The proposed method can also be suitable for clinical intention after successful pre-training a neural network by specific patterns of the different network topologies.

### Acknowledgement

This work was supported by the Iran Neural Technology Research Centre, Tehran, Iran under Grant 48.M.111194.